\documentclass{PoS}

\title{Two pion scattering, $I=0$ and disconnected diagrams in 
Lattice QCD}

\ShortTitle{Disconnected diagrams}

\author{\speaker{Qi Liu}\\
        Department of Physics, Columbia University, New York, NY 10025, USA \\
        E-mail: \email{qiliu@phys.columbia.edu}}

\author{RBC and UKQCD collaborations}

\abstract{A full, high-statistic $I=0$ and $I=2$ $\pi-\pi$ scattering calculation including the disconnected diagram is performed on 2+1 flavor, domain wall fermion gauge configurations generated using the Iwasaki gauge action with $\beta=2.13$, a $16^3\times32$ volume with $L_s=16$ and $a^{-1}=1.73GeV$. Using the same light quark propagators and additional strange quark propagators, we study the $\eta$ and $\eta^\prime$ mesons, where disconnected diagrams also make an important contribution. Our $\pi-\pi$ calculation shows a good exponentially decaying signal from the disconnected graph, which makes a small contribution to the $I=0$ scattering amplitude at least for the short times where it can be accurately computed. We are able to resolve the $\eta$ and $\eta^\prime$ states and see the pattern of SU(3) flavor symmetry breaking found in Nature. }

\FullConference{The XXVII International Symposium on Lattice Field Theory - LAT2009\\
		 July 26-31 2009\\
		 Peking University, Beijing, China}

\begin{document}

\section{Introduction}
Disconnected diagrams play a very important role in many lattice QCD calculations, such as the calculation of $\eta$, $\eta'$, and $\sigma$ masses, the isospin 0 $\pi-\pi$ scattering length, $\epsilon'$ from K to $\pi\pi$ $\Delta I=1/2$ process, \textit{etc}.. Since the correlation functions corresponding to these diagrams reflect the quantum fluctuations of QCD vacuum, the noise does not exponentially decrease with increasing time separation between source and sink. Thus, the signal present in these correlation functions (which behaves like $\exp{(-m_{eff}t)}$ for large time separations) is quickly buried in the noise and it is very difficult to get useful information from large time separation. Therefore, a good understanding of the disconnected diagrams requires a full dynamical calculation and large statistics. Several attempts related to these calculation have been done for the $\eta'$ in the case of SU(2)~\cite{Hashimoto:2008p3440, Jansen:2008p2831, Michael:2001p2344} and the $\sigma$ ~\cite{Mathur:2007p3592, Kunihiro:2008p3239}. It would be more interesting to calculate the physical $\eta$ and $\eta'$ masses in the flavor symmetry broken SU(3) case where $m_s \textgreater m_l$.

The $\pi-\pi$ scattering length calculation per se, is very important to expand our understanding of the strong interaction on lattice QCD. It also represents a first step toward a more interesting calculation of the direct CP violation measure $\epsilon'$ which is believed to be a quantity that may call for new physics.  While much work has been dedicated to the $I=2$ channel of $\pi-\pi$ scattering~\cite{Aoki:2002p2783, Meng:2004p2785}, few lattice QCD calculations have been done for the $I=0$ case, possibly because the disconnected diagram makes the problem much harder than $I=2$.  Y. Kuramashi \textit{et.~al.~}studied the isospin 0 channel $\pi-\pi$ scattering~\cite{KURAMASHI:1993p2591} almost two decades ago, but he ignored the disconnect diagram and also used the quenched approximation. In this paper, we calculate all the diagrams, especially providing a close look at the disconnect diagram. We will get a rough idea about how large the disconnected diagram's contribution is. 

\section{Details of Lattice Calculation}
Our calculation is a full unitary calculation and is based on the RBC/UKQCD $16^3\times32$, $L_s=16$, 2+1 flavor domain wall fermion, $\beta=2.13$ Iwasaki gauge action lattices. The inverse lattice spacing for these lattices is determined to 1.73(3)GeV~\cite{Allton:2008p573}. We have three ensembles of such configurations with light sea quark mass $m_l$=0.01, 0.02, 0.03 respectively. They all have the same strange sea quark mass $m_s=0.04$. The ensembles with $m_l=0.01$ and 0.02 each has 150 configurations, and the $m_l=0.03$ ensemble has 281 configurations. 

We use a wall source and a wall sink for the propagator. The calculation of the correlation function for the disconnected diagrams naturally requires us to calculate the propagators on the time slices of both source and sink, so we decide to calculate the propagators on all time slices, which requires the calculation of 32 separate propagators in our case. By doing so, the statistics are greatly improved since we can put the source at all possible time slices. In a sense, we extract all possible information from a given configuration. Finally, we exploit the need to compute many propagators on each configuration by using Ran Zhou's eigenvector accelerator code, speeding up the computation of the propagators by 60 percent. 

\section{The $\eta$ and $\eta\prime$ masses}
\begin{figure}[!tb]
	\begin{minipage}{0.5\textwidth}
	\begin{tabular}{cc}
	\includegraphics[width=0.45\textwidth]{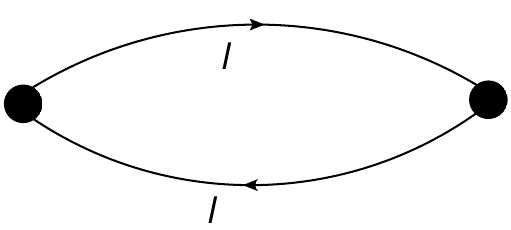} &
	\includegraphics[width=0.45\textwidth]{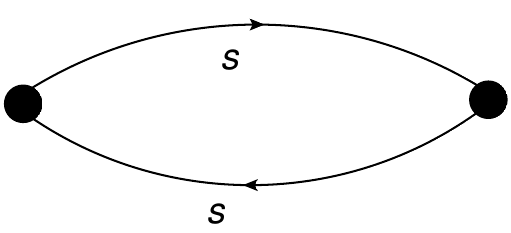}\\
	\includegraphics[width=0.45\textwidth]{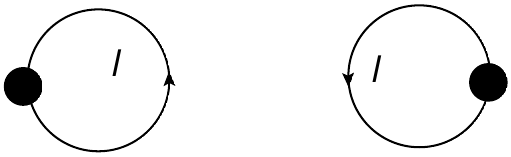} &
	\includegraphics[width=0.45\textwidth]{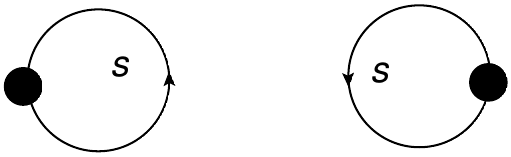}\\
	\includegraphics[width=0.45\textwidth]{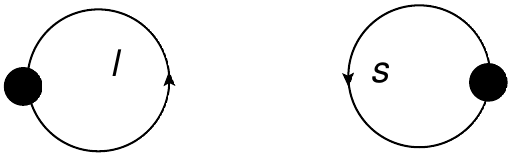} &
	 \\
	\end{tabular}
	\end{minipage}
	\begin{minipage}{0.5\textwidth}
	\begin{center}
	\includegraphics[width=0.9\textwidth]{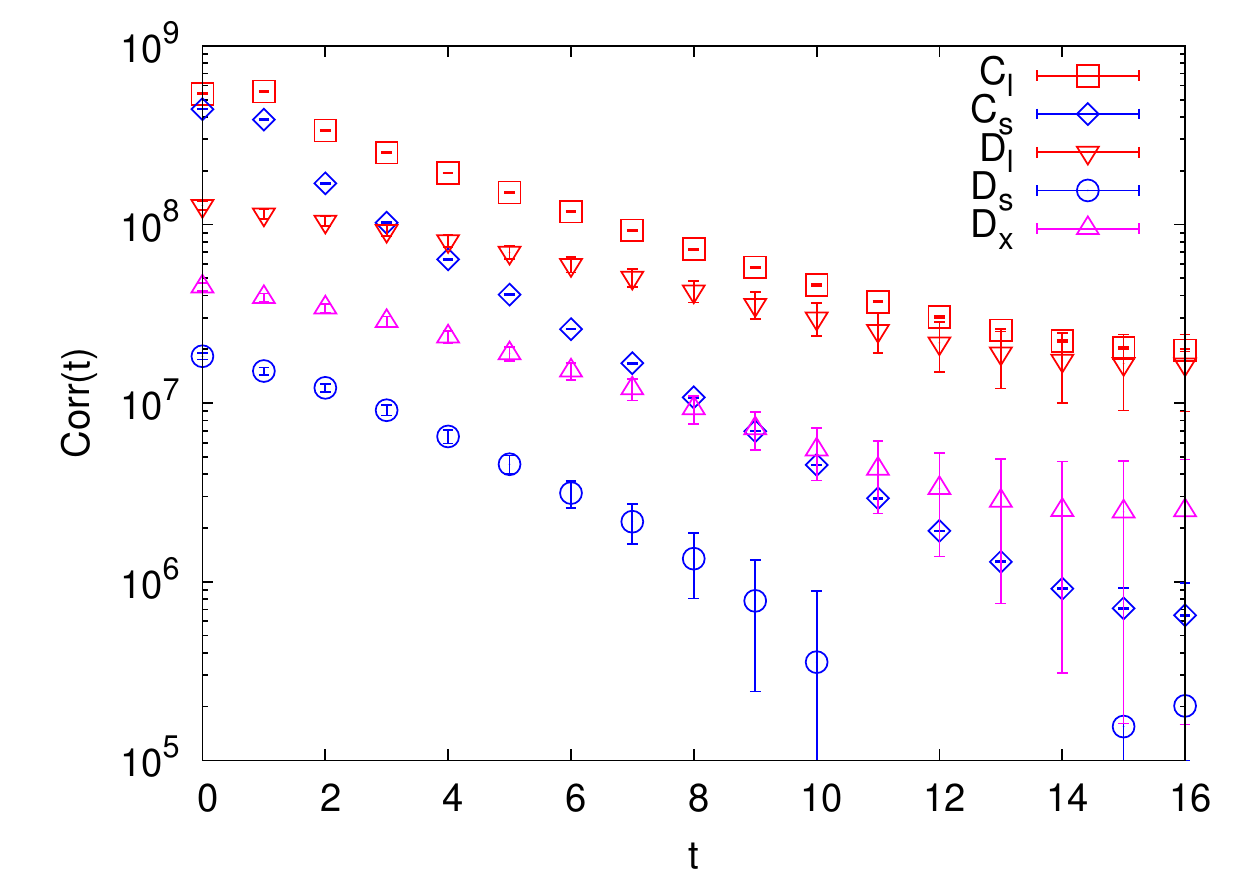}
	\end{center}
	\end{minipage}
	\caption{The left panel shows all five diagrams that contribute to the $\eta$ and $\eta'$ correlation functions. They are $C_l$, $C_s$, $D_l$, $D_s$, and $D_x$ respectively from the top left to bottom right. The solid line with an arrow stands for a propagator, and the black solid circle stands for a $\gamma_5$ insertion. The right panel shows our result for these diagrams for the $m_l=0.01$ ensemble.}
	\label{fig:pseudoScalarAll}
	\end{figure}

Using approximate SU(3) flavor symmetry, we expect that we can create the $\eta$ with the octet operator $O_8$, and the $\eta'$ with the singlet operator $O_1$,
	\begin{eqnarray}
	\eta & \approx O_8=&\frac{1}{\sqrt{6}}(\bar{u}\gamma_5u+\bar{d}\gamma_5d -2\bar{s} \gamma_5s)  \label{eqn:O8} \\
	\eta' & \approx O_1=&\frac{1}{\sqrt{3}}(\bar{u}\gamma_5u+\bar{d}\gamma_5d +\bar{s} \gamma_5s) \label{eqn:O1}
	\end{eqnarray}
We will try to demonstrate this is true from lattice QCD calculation. For future convenience, we define a light operator and a strange operator,
	\begin{eqnarray}
	O_l &=& \frac{1}{\sqrt{2}}(\bar{u}\gamma_5u+\bar{d}\gamma_5d) \\
	O_s &=& \bar{s}\gamma_5s
	\end{eqnarray}
and consider the correlation function matrix of these two operators,
	\begin{eqnarray}
	C(t)=\left(\begin{array}{cc}
	\langle O_l(t)  O_l^\dagger(0) \rangle & \langle O_s(t)  O_l^\dagger(0) \rangle \\ 
	\langle O_l(t)  O_s^\dagger(0)  \rangle & \langle O_s(t)  O_s^\dagger(0) \rangle  
	\end{array}\right)=\left(\begin{array}{cc}
	C_l-2D_l & -\sqrt{2}D_x \\
	-\sqrt{2}D_x & C_s-D_s
	\end{array}\right)
	\label{eqn:OpMatrix}
	\end{eqnarray}
where $C_l$, $C_s$, $D_l$, $D_s$, and $D_x$ stand for all the five possible contractions which are shown in Figure~\ref{fig:pseudoScalarAll}. 
Then by diagonalizing the matrix $C^{-1}(t_0)C(t)$ with a reference time $t_0=1$, we expect to get the correlation function for $\eta$ \& $\eta'$ states, while the rotation matrix that diagonalizes this matrix gives the mixing angle for the light and the strange parts,
	\begin{eqnarray}
	\eta & = & \cos\theta_\eta \cdot O_l - \sin\theta_\eta\cdot O_s \\
	\eta' &=& \sin\theta_{\eta'} \cdot O_l + \cos\theta_{\eta'}\cdot O_s 
	\end{eqnarray}
In the SU(3) flavor symmetry limit, $\sin\theta_{\eta,\eta'}=\sqrt{\frac{2}{3}}=0.8165$, $\cos\theta_{\eta,\eta'}=\sqrt{\frac{1}{3}}=0.5774$. 

\begin{figure}[!tb]
	\begin{tabular}{ll}
	\includegraphics[width=0.45\textwidth]{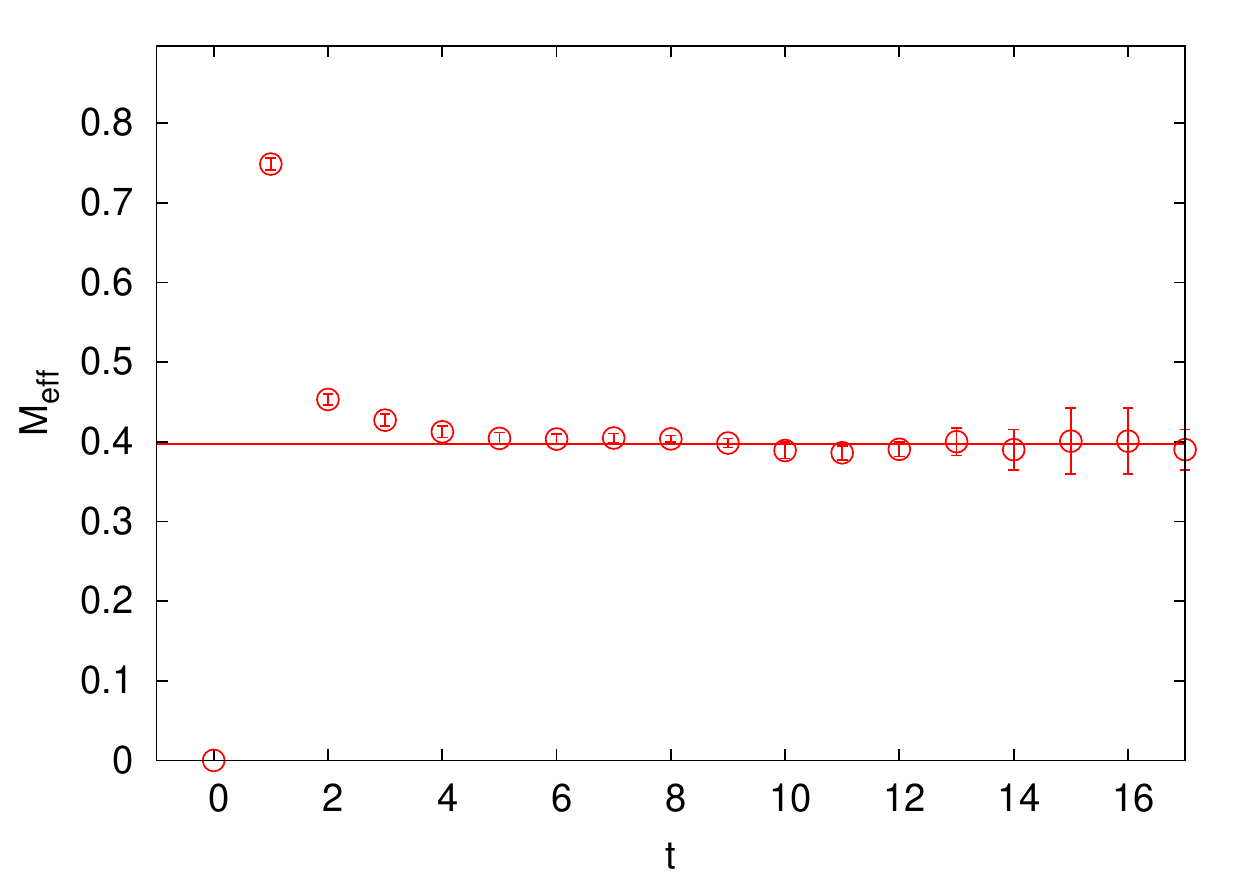} &
	\includegraphics[width=0.45\textwidth]{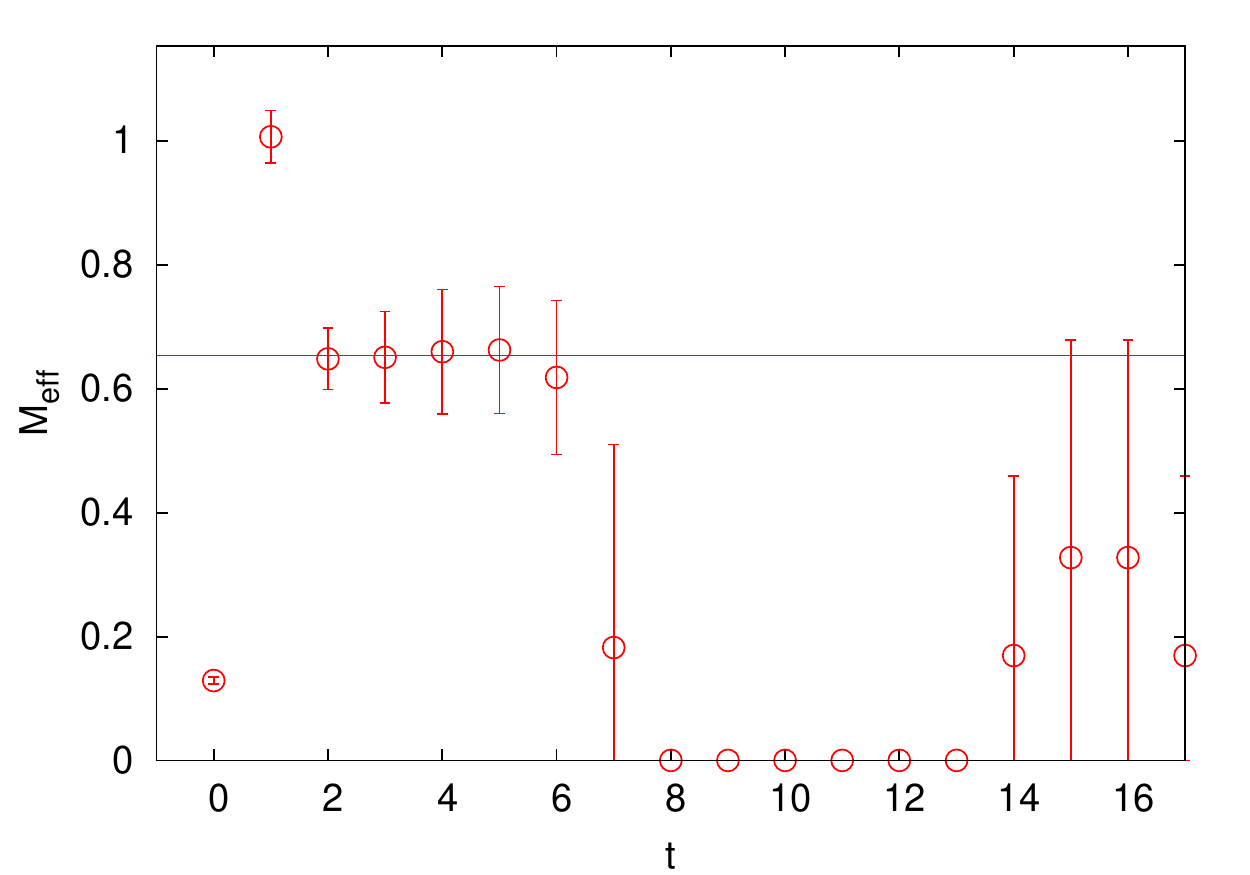} \\
	\end{tabular}
	\caption{Left panel: effective mass plateau for the $\eta$ and the mass obtained using the fitting range [5:14]. Right panel: effective mass plateau for the $\eta'$ and the mass obtained using the fitting range [3:7]. }
	\label{fig:meffEtaEtap}
	\end{figure}

\begin{figure}
	\begin{tabular}{cc}
	\includegraphics[width=0.5\textwidth]{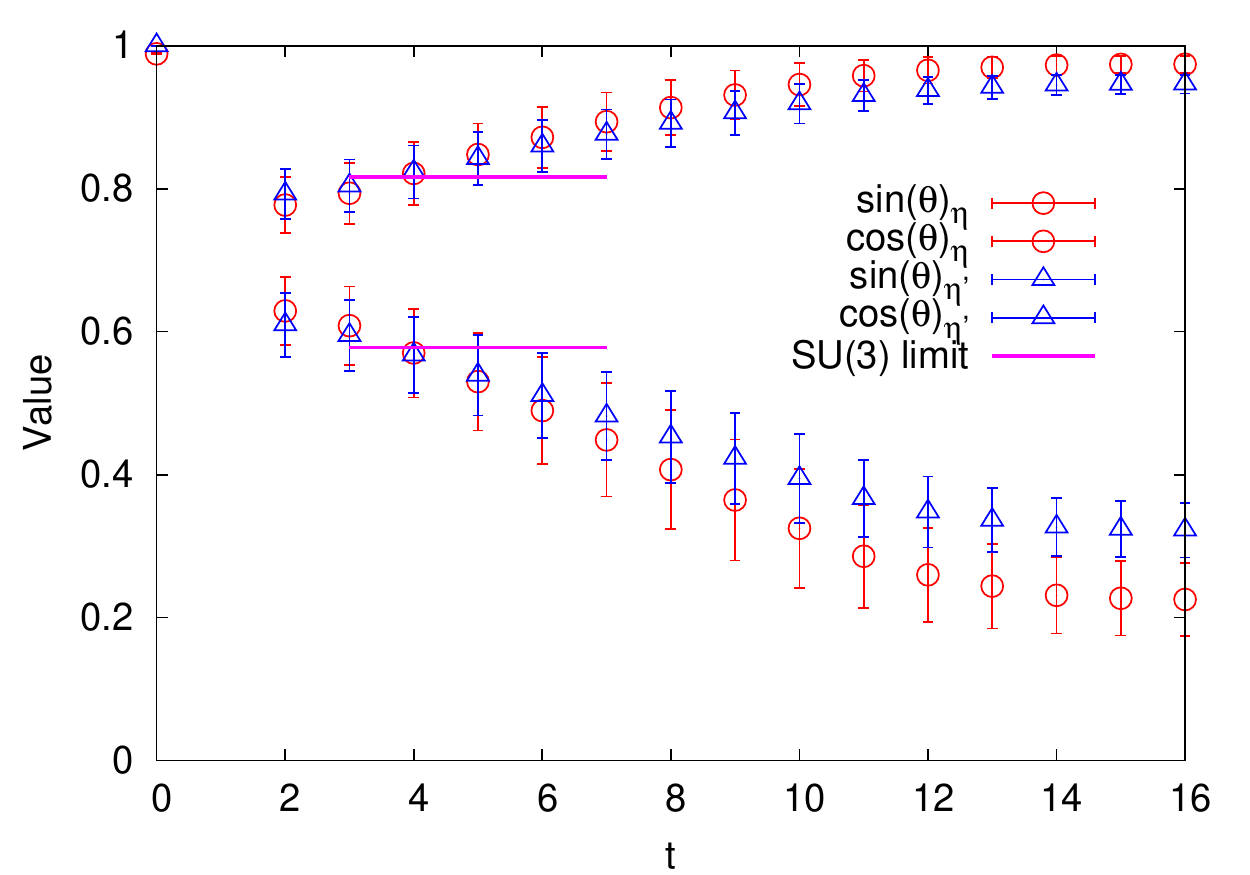} &
	\includegraphics[width=0.5\textwidth]{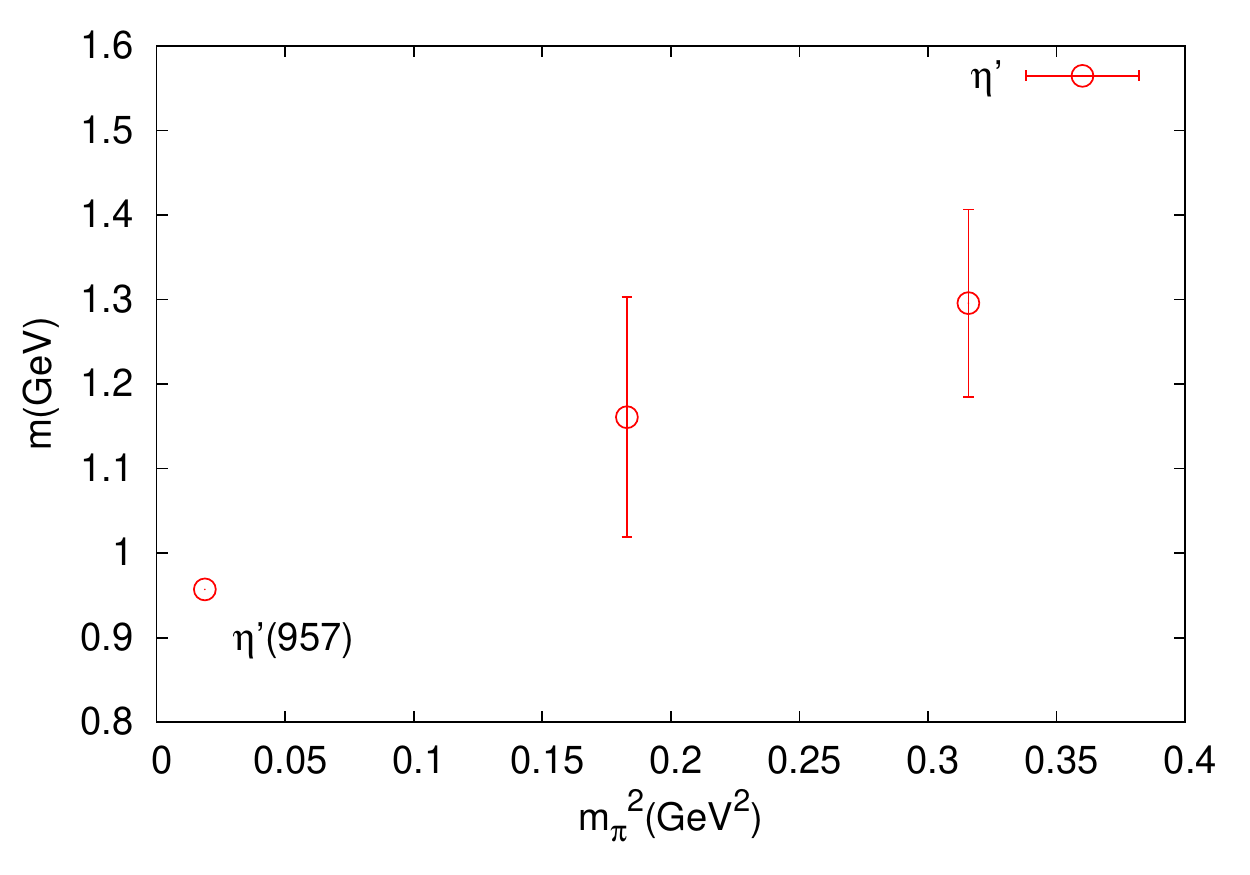} \\
	(a) & (b)
	\end{tabular}
	\caption{(a). The Mixing angle, which determines the two eigenvectors of the matrix $C^{-1}(t_0)C(t)$. The two horizontal lines give the value of $\cos(\theta)$ and $\sin(\theta)$ in the SU(3) flavor symmetry limit. (b). The $\eta'$ mass from the $m_l=0.01$ and $m_l=0.02$ ensembles. The physical mass 957MeV is also labeled on the figure. }
	\label{fig:mixAngleEtap}
	\end{figure}

The effective mass plateau of $\eta$ and $\eta'$ for $m_l=0.01$ ensemble are shown in Figure~\ref{fig:meffEtaEtap}. Using the range $3\leq t \leq 7$ where there are good plateaus,  we get $m_\eta=0.397(17)=687(29)MeV$, and $m_{\eta'}=0.654(80)=1.13(14)GeV$. Figure~\ref{fig:mixAngleEtap}(a) shows the mixing angle calculated by diagonalizing the correlation matrix. The data points with small time separation where we have good signal for the $\eta'$ agree with the SU(3) symmetry limit. This demonstrates that the $\eta$ is created by the octet operator and the $\eta'$ by the singlet operator, approximately. This ability of a full QCD calculation to reproduce the anomaly dominated SU(3) symmetry breaking pattern of the physical $\eta$ and $\eta'$ is an important test of lattice QCD.

	Since the SU(3) flavor symmetry is approximately true, using Eqn.~\ref{eqn:O8} and Eqn.~\ref{eqn:O1}, we can calculate $\eta$ and $\eta'$ correlation function directly from the $O_8$ and $O_1$ operator,
	\begin{eqnarray}
	\langle\eta(t) \eta^\dagger(0)\rangle &=& \frac{1}{3}(C_l+2C_s)-\frac{2}{3}(D_l+D_s-2D_x)\\
	\langle\eta'(t)\eta'^\dagger(0)\rangle & = & \frac{1}{3}(2C_l+C_s)-\frac{1}{3}(4D_l+D_s+4D_x)
	\end{eqnarray}
	this gives $ m_\eta=0.392(12) = 678(20) MeV $,  $ m_{\eta'}=0.671(82) = 1.16(14)GeV $, and they agree with the results of correlation matrix diagonalization method. 
	
	Our result for the $\eta$ mass is consistent with Gell-Mann-Okubo formula, which states that $3m_\eta^2+m_\pi^2=4m_k^2$, and gives $ m_\eta=0.3855(25)=667(4)MeV $ , where we have used $m_\pi=0.2472(10)=427.7MeV$ and $m_k=0.3560(20)= 615.9MeV$.
	
	For the $\eta'$ mass, since we can only get a reasonable signal from short time separation where the excited states may exist, our result may not be fully reliable. Figure~\ref{fig:mixAngleEtap}(b) shows $m_{\eta'}$ calculated for $m_l=0.01$ and 0.02 ensembles. By assuming a linear relationship between $m_{\eta'}$ and $m_{\pi^2}$, it seems our result does agree with the experiment physical $\eta'$ mass.
		
\section{Two-Pion Scattering}
\begin{figure}
	\begin{minipage}{0.45\textwidth}
	\begin{tabular}[c]{ll}
	\includegraphics[width=0.45\textwidth]{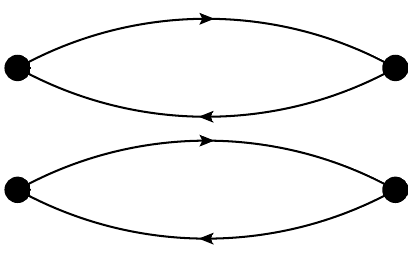}&
	\includegraphics[width=0.45\textwidth]{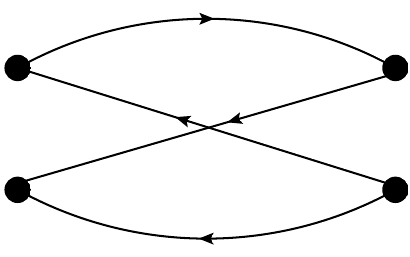}\\
	\includegraphics[width=0.45\textwidth]{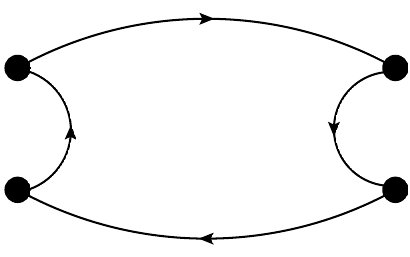}&
	\includegraphics[width=0.45\textwidth]{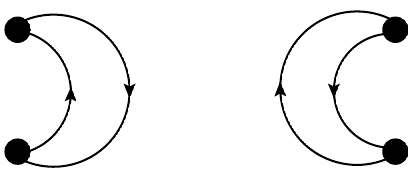}
	\end{tabular}
	\end{minipage}
	\begin{minipage}{0.5\textwidth}
	 \includegraphics[width=\textwidth]{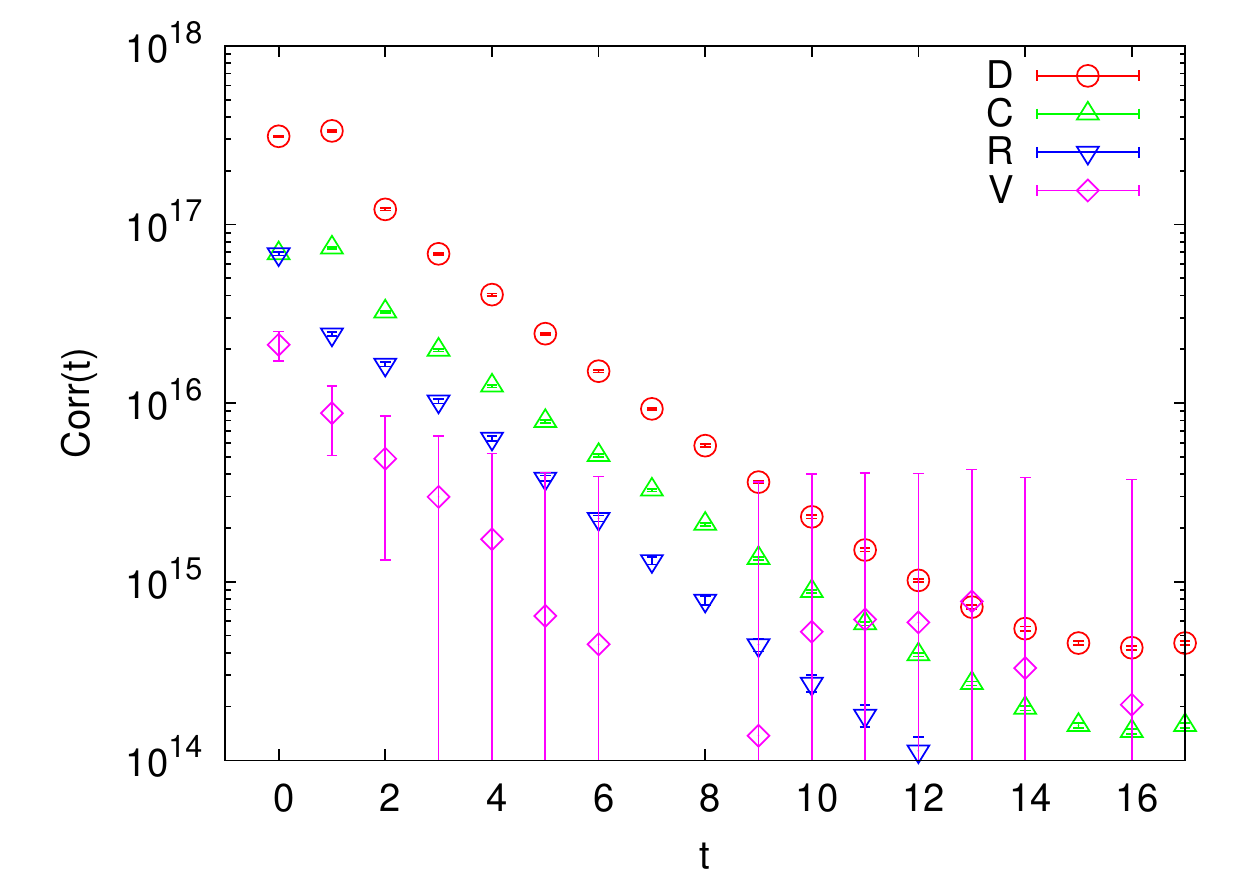}
	 \end{minipage}
	 \caption{The left panel shows the Direct(D), Cross(C), Rectangular(R), and Vacuum(V) diagrams from the left top to right bottom. The right panel shows our result for these diagrams for the $m_l=0.01$ ensemble.}
	 \label{fig:DCRV}
	 \end{figure}

	The diagrams that need to be calculated in order to study $\pi-\pi$ scattering in both $I=2$ and $I=0$ channels are shown in the left panel of Figure~\ref{fig:DCRV}, identified as the Direct(D), Cross(C), Rectangular(R), and Vacuum(V) diagrams. The right panel shows the actual data for these four contractions for the $m_l=0.01$ ensemble. We can see that diagram D makes the biggest contribution, then diagram C and R, and diagram V makes the smallest contribution. It is extremely noisy for the disconnected diagram(V), but still we can get a clear signal up to time separation 6. By increasing the statistics, or reducing the pion mass(making the $I=0$ signal fall less rapidly), we could obtain a better result for the V diagram. 
	
	The $\pi-\pi$ correlation function for $I=2$ and $I=0$ can be expressed in terms of these diagrams,
	\begin{eqnarray}
	 <I_2(t)|I_2(0)> &=& 2(D-C)\\
	 <I_0(t)|I_0(0)> &=& 2D+C-6R+3V
	 \end{eqnarray}
We can get the energy of $I=2$ and $I=0$ $\pi-\pi$ state from the plateau of the effective mass plot, and the results are shown in Table~\ref{tab:energy}. Since the disconnected diagram makes only a small contribution to the correlation function of $I=0$ scattering state but contributes very big noise, leaving it out makes the numerical result more precise but in trade with an unknown systematic error, and it is listed as $\Delta_E(I=0$ Vout) in the table. 
	 
	\begin{table}
	\caption{The energy difference between the $\pi-\pi$ isospin eigenstates and 2$m_\pi$. The last column shows $\Delta_E(I=0)$ when we leave the disconnected diagram out. The numbers in the bracket [] give the fitting range.}
	\begin{center}
	\begin{tabular}{ccccc}
	\hline
	\hline
	$m_l(conf)$ & $m_\pi/[5:14]$ & $\Delta_E(I=2)/[5:14]$ & $\Delta_E(I=0)/[5:8]$ & $\Delta_E(I=0$ Vout)/[5:12] \\
	\hline
	0.01(150) & 0.2472(10) & 0.0197(14) & -0.041(80) & -0.0439(72) \\
	0.02(150) & 0.3248(9) & 0.0158(7) & -0.086(79) & -0.0268(69) \\
	0.03(281) & 0.3895(6) & 0.0135(5) & 0.04(10) & -0.0155(50)\\
	\hline	
	\end{tabular}
	\end{center}
	\label{tab:energy}
	\end{table}
	
	\begin{table}
	\caption{The $\pi-\pi$ scattering lengths for the $I=2$ and $I=0$ channels. The data in the bracket [] shows the minimum and maximum value corresponding to one standard deviation. The last column shows the result for $I=0$ scattering length when the disconnected diagram is omitted.}
	\begin{center}
	\begin{tabular}{cccc}
	\hline
	\hline
	$m_\pi$ & $a_0(I=2)$ & $a_0(I=0)$ & $a_0(I=0$ Vout) \\
	\hline
	0.2472(10) & -1.26(7) & 4.6 [-2.1:8.1] & 4.9 [4.2:5.4] \\
	0.3248(9) & -1.31(5) & 7.9 [0.86:10.1] & 4.1 [3.0:4.9] \\
	0.3895(6) & -1.34(5) & -2.9 [-6.0:7.3] & 2.8 [1.7:3.8] \\
	\hline
	\end{tabular}
	\end{center}
	\label{tab:scaLen}

	\end{table}
	
\begin{figure}[!tb]
	\centering
	\includegraphics[width=0.7\textwidth]{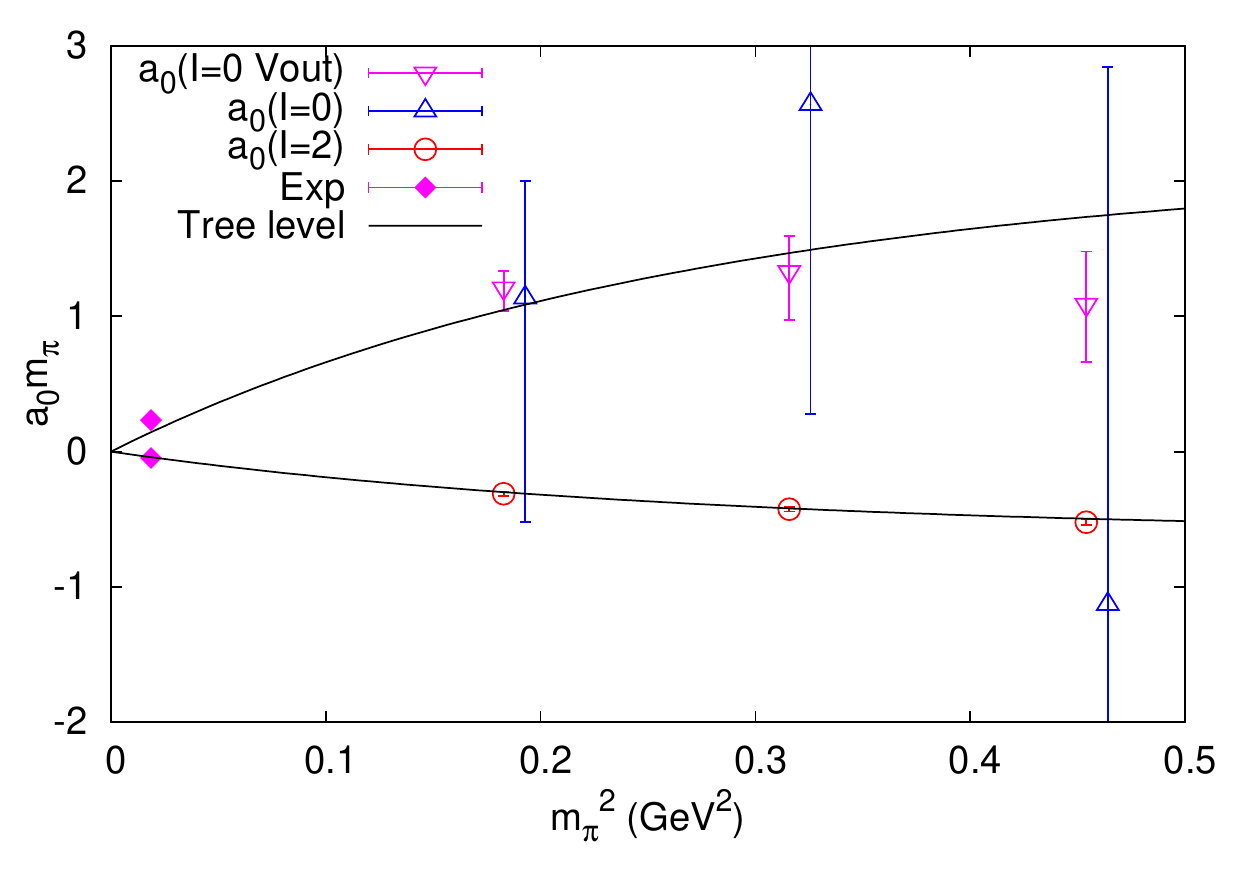}
	\caption{The $\pi-\pi$ scattering length for $I=2$ and $I=0$. The solid line gives the tree level prediction.}
	\label{fig:scaLen}
	\end{figure}

	To get the scattering length, we use L\"{u}scher formula~\cite{Luscher:1991p2480, Lellouch:2001p4241}, which relates the energy of the two particle states with the scattering length,
	\begin{eqnarray}
	\Delta E = E -  2m_\pi = -\frac{4\pi a_0}{m_\pi L^3}[1+c_1 \frac{a_0}{L}+c_2(\frac{a_0}{L})^2)]+O(1/L^6)
	\end{eqnarray}
where $c_1$=-2.8373, $c_2$=6.3752. The calculated scattering lengths are shown in Table~\ref{tab:scaLen} and Figure~\ref{fig:scaLen}.

	 Let us compare our result with theory preditions. The tree level $\chi$PT formula for the $\pi-\pi$ scattering length is
	 \begin{equation}
	a_0(I=2)=-\frac{1}{16\pi}\frac{2m_\pi^2}{f_\pi^2} , a_0(I=0)=\frac{1}{16\pi}\frac{7m_\pi^2}{f_\pi^2} 	
	\end{equation}
	which we evaluate using the results for $f_\pi$ and $m_\pi$ obtained in Ref~\cite{Allton:2007p2751},
	\begin{equation}
	f_\pi=0.0765+1.02(m_l+0.00308), m_\pi^2=2\times 2.285(m_l+0.00308)
	\end{equation}
The $\chi$PT results are shown in Figure~\ref{fig:scaLen} as the solid line, in comparison with our lattice QCD results. Notice that the tree level formula agrees well with the experiment and our results for the $I=2$ scattering length, while giving a value for $I=0$ around 30 percent smaller compared to experiment(see Figure 1 in Ref~\cite{Leutwyler:2006p3178}). 	

\section{Conclusion}
We have preformed a unitary, full lattice QCD calculation of the $\eta$ and $\eta'$ mass, and the isospin 0 $\pi-\pi$ scattering length where the disconnected graph plays an important role. The $\eta$ mass we obtain is in a good agreement with chiral pertubation theory. We still have large errors for the $\eta'$ mass, and the full $I=0$ $\pi-\pi$ scattering length has such a big error that the result serves only as a bound on the magnitude . A clear signal can only be seen for small time separation in the disconnected diagram of $\pi-\pi$ scattering.

In the future, increasing the statistics alone could improve our results to some extent. Reducing the vacuum noise by doing the calculation on a larger volume or making the pion mass smaller could greatly improve the signal to noise ratio of the vacuum diagram. As more powerful computers are under construction, we will make the calculation of the disconnected diagram more precise, and hope to attack the $\epsilon'$ problem.

{\bf Acknowledgements}
I thank all my colleagues in the RBC and UKQCD collaborations for discussions, suggestions, and help with program development. I especially thank Norman Christ for detailed instructions, and Ran Zhou for providing the eigenvector accelerator code. I acknowledge Columbia University, RIKEN, BNL and the U.S. DOE for providing the facilities on which this work was performed. This work was supported in part by U.S. DOE grant \#DE-FG02-92ER40699.

\end{document}